\documentclass{article}

\usepackage{arxiv}

\usepackage[utf8]{inputenc} 
\usepackage[T1]{fontenc}    
\usepackage{hyperref}       
\usepackage{url}            
\usepackage{booktabs}       
\usepackage{amsfonts}       

\usepackage{microtype}      
\usepackage{lipsum}
\usepackage{graphicx}
\usepackage{longtable}
\usepackage{rotating}

\usepackage{tikz}

\usepackage{colortbl}
\graphicspath{ {./images/} }

\title{Low-Cost SMS Driven Location Tracking Platform Towards Anti-Poaching Efforts}

\author{
  Jack Burkett \\
  School of Computer Science and Informatics\\
  Cardiff University, UK \\
  \texttt{BurkettJJ@cardiff.ac.uk} 
  \And
  Pablo Orozco Ter Wengel \\
  School of Biosciences \\
  Cardiff University, UK \\
  \texttt{orozco-terwengelpa@cardiff.ac.uk} 
  \And
  Benoit Goossens \\
  School of Biosciences \\
  Cardiff University, UK \\
  \texttt{goossensbr@cardiff.ac.uk} 
  \And
  Omer Rana \\
  School of Computer Science and Informatics\\
  Cardiff University, UK \\
  \texttt{ranaof@cardiff.ac.uk} 
  \And 
  Charith Perera \\
  School of Computer Science and Informatics\\
  Cardiff University, UK \\
  \texttt{pererac@cardiff.ac.uk} 
}
\usepackage{subcaption}

\begin{document}
\maketitle
\begin{abstract}
Throughout the world, poaching has been an ever-present threat to a vast array of species for over many decades. Traditional anti-poaching initiatives target catching the poachers. However, the challenge is far more complicated than catching individual poachers. Poaching is an industry which needs to be fully investigated. Many stakeholders are directly and indirectly involved in poaching activities (e.g., some local restaurants illegally providing meat to tourists). Therefore, stopping or severely decapitating the poaching industry requires a unified understanding of all stakeholders. The best way to uncover these geographical and social relationships is to track the movements of poachers. However, location tracking is challenging in most rural areas where wildlife sanctuaries are typically located. Internet-connected communication (e.g. 3G) technologies typically used in urban cities are not feasible in these rural areas. Therefore, we decided to develop an SMS (short message service) base low-cost tracking system (SMS-TRACCAR) to track poachers. The proposed system was developed to be deployed in Kinabatangan Wildlife Sanctuary, Sabah, Malaysia and nearby villages and cities where poachers typically move around. Our evaluations demonstrated that SMS-based tracking could provide sufficient quality (granular) data (with minimum energy consumption) that enable us to monitor poacher vehicle movements within rural areas where no other modern communication technologies are feasible to use. However, it is important to note that our system can be used in any domain that requires SMS-based geo-location tracking. SMS-TRACCAR can be configured to track individuals as well as groups. Therefore, SMS-TRACCAR contributes not only to the wildlife domain but in the wider context as well.

\end{abstract}



\maketitle
\newpage

\section{Introduction}

\begin{figure*}[h!]
	\centering
	\includegraphics[scale = 1.0]{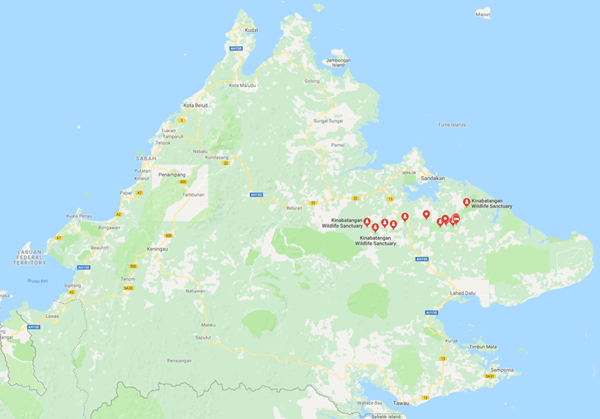}
	\caption{A map of the Kinabatangan Wildlife Sanctuary. Sandakan is the closest major city.}  
	\label{Fig:KinabatanganWildlifeSanctuary}
\end{figure*}

Sabah \cite{Ancrenaz2012, PereraTR2018}, a Northern state in Malaysia, is championed for its extensive biodiversity that thrives off the equatorial climate and tropical rainforests. With such large biodiversity, it is no wonder ecotourism is one of the major industries of the state \cite{Chan2007} and many of Sabah’s visitors are destined for Kinabatangan, a region of Sabah that has been described as ‘Sabah’s Gift to the Earth’ \cite{WWF191}. However, this breadth of species in this area is the consequence of a number of economic activities around the region that have had a detrimental impact on wildlife. Logging activities and development of areas for agriculture for crops such as coffee, cocoa, rubber, tobacco and in most recent times, palm oil have pushed the species of animals and plants into a thin corridor known as the corridor of life. With such a vast array of species in a relatively small area, it means that it has become a very lucrative spot for poachers. Even though in 2005, under the state’s Wildlife Conservation Enactment of 1997, a total of 26,000ha was gazetted as the Kinabatangan Wildlife Sanctuary, it has done little to prevent the prevalent poaching in the region \cite{WWF191}. Whilst the Sumatran Rhinoceros has been extinct in Malaysia since 2015 \cite{WWF191}, the horns of the Rhinoceros can fetch as much as \$30,000 per kilogramme \cite{Chan2017} demonstrating how profitable poaching can be. It also highlights a clear demand for these products, which is a huge part of the problem with poaching, especially in Sabah. Alarmingly, a 2010 TRAFFIC report on the Pangolin trade in Sabah found that 22,200 Pangolins were traded in 13 months by a trade syndicate \cite{Ale13}. A map of the Kinabatangan Wildlife Sanctuary is presented in Figure \ref{Fig:KinabatanganWildlifeSanctuary}.

\textbf{Contribution:} The main aim of this project was to create a low-cost system that uses cheaper consumer-grade location trackers and allows the user to track the location of vehicles using modified open-source GPS software without requiring internet access. Further, the devices must be able to track vehicles over a sufficient period of time, for example, a minimum of two to three days. The system should also be easy to use and require minimal configuration from the user to work and operate. We proposed SMS-TRACCAR to address these requirements.

However, it is important to note that our system can be used in any domain that requires SMS-based geo-location tracking. SMS-TRACCAR can be configured to track individuals as well as groups. For example, our solution can be re-utilised to track bicycles in a smart city or joggers in the city without any changes (as groups of individuals). Therefore, SMS-TRACCAR contributes not only to the wildlife domain but in the wider context as well.

\section{Background and Motivation}

In this section, we briefly discuss different approaches that have been proposed to track vehicles.

\subsection{Vehicle Tracking using RFID/BLE/Vision Technology}
In order to track vehicles even in places such as tunnels, Ning et al. \cite{YNi13} have proposed a method of using RFID, GPS, GPRS and the LANDMARC method. They recognised that using solely GPS has shortcomings in terms of accuracy and coverage and that many of the previous solutions do not provide a method of acquiring real-time information from the vehicle in areas such as tunnels or in narrow streets surrounded by high-rise buildings. To address this issue, they proposed the use of RFID along with the LANDMARC approach, which is an algorithm that originates from an 'active RFID indoor positioning System' and allowed them to create a solution that does not rely on expensive RFID readers. Along with an RFID module, the solution also contains a GPRS, GPS and sensing module that are used together to form a vehicle controller, which is the core of the location system. As the car drives along, the RFID reader reads the tags on the road, allowing the system to know the vehicle's real-time position accurately. After testing their solution, they found that, under complex environments, the system successfully located the vehicle's position. However, in the future, further attention would have to be paid to the stability and reliability of the system. Whilst utilising RFID and other sensor technologies can improve the accuracy of a location system, as demonstrated by this solution, the problem with them is that it involves the placement 5of many tags throughout a location. This is time-consuming to configure and very expensive to deploy, depending on the size of the location. As this system proposed in this paper would be placed in cities/towns, it makes the use of RFID in this way infeasible. For this reason, we decided to use GPS technology without extra sensors to ensure the system is cost-effective when deploying multiple devices in a location. Other approaches have also been developed to track vehicles using Bluetooth Low Energy (BLE) technologies \cite{Karan2020}. Due to the short-range nature of both RFID and BLE, they require a large network of sensor nodes to track vehicles. However, BLE is useful to see whether a particular vehicle has visited a particular critical location (e.g. usual known spot where poacher stops their vehicles). Another method has been to use computer vision techniques to detect licence plate numbers. There are many challenges \cite{Shashirangana2021b} to detecting vehicle licence plates covertly in remote jungle areas, such as limited visibility and lighting \cite{Padmasiri2022}, limited connectivity, limited access to computation and energy sources \cite{Shashirangana2021}, exposure to harsh jungle environments (e.g. humidity, water etc.) and so on. However, vision technologies can be used to augment the other solutions and deploy them in strategic locations where road access is restricted to a certain time period of a day or certain authorised vehicles.

\subsection{Vehicle Tracking using GPS/GSM/GPRS Technology}

With a focus on a solution that solely uses GPS to track vehicles, \cite{Raj11} proposed a system that uses GPS and GPRS on cheap mobile phone devices to provide real-time positioning of the phone. They recognised that many commercial location devices available in Mauritius are rather expensive and do not exploit the use of mobile devices for GPS tracking. They also found that whilst mobile phone users can download location tracking applications on their phones, many of these have limited functionality or require paying extra for some features. The system proposed works on any J2ME and internet-enabled mobile phone and uses a Java Midlet application installed on the phone to retrieve the location from the phone's GPS and sends it to the server using GPRS. After testing the system, they found that over a track of 50 kilometres, the proposed system was cheaper to run and matched the accuracy of the cheapest commercial device available. Even though the system was successful, it is only compatible with phones with MIDP 2.0 (Mobile Information Device Profile), and the short span of the battery life of the phone is a factor that must be considered. The main lesson we learnt was that if a given system were to use GPRS, it makes it less cost-effective due to the costs associated with GPRS data packages. GPRS also consumes a significant amount of energy. In contrast, our approach to using SMS is much cheaper in terms of both energy consumption and running costs.

\subsection{Vehicle tracking using GPS/GSM Technology}

Using only GPS and GSM technology to track vehicles is a more cost-effective method and more viable in locations with a lack of infrastructure than using GPRS. Pham et al. \cite{Pha13} have proposed a system that uses a hardware prototype with a GPS and GSM module in order to track vehicles. They found that current vehicle tracking options were often application and region-specific as well as costly and, therefore, proposed this system as a cheaper, more versatile alternative. The prototype used consists of an Arduino with a GPS receiver and a GSM module attached. The prototype works by retrieving the current location using the GPS module and sending the coordinates to a receiving mobile phone device as SMS messages. After testing the system, they found that its accuracy was as good as the commercial devices whilst also providing greater customisability and global operability at a lower cost. This result influenced our solution.

\subsection{Tracking animals using GPS/GSM Collars}

Ericsson et al. \cite{Hol04} have proposed a system that uses GPS/GSM collars to track moose in Sweden. This system was proposed in response to the problem of many previous techniques that used UHF (Ultra High Frequency) or VHF (Very High Frequency) radio collars requiring significant effort in the field to collect data. The system proposed works by sending the location of the collar, which is attached to the moose, via SMS to a modem. The modem is connected to a database which stores the location data which can then be viewed on an interactive map through a web application. After testing this system over a period of four years they found it works and gives a new possibility to ‘supervise’ animals in almost real-time. They also found the biggest disadvantage was the limited or non-existing coverage of GSM networks in remote areas in the region, however they also note that less GSM coverage is required for SMS transmissions in comparison to making a telephone call or GSM data transfer. The second paper Evans et al. \cite{Hol04} have proposed to provide a method of tracking Malay Civets in Sabah, Malaysia without using bulky collar units. The collars used in the proposed system utilise a GPS microchip, a 2300 or 2500 mAh battery, a UHF radio transmitter, a tri-axial accelerometer and an antenna. Following testing of the system they found that the system returned a GPS success of 58.1\% across all tracked individuals over the testing period, which was up to 187 days long. Both of these works demonstrate that GPS and GSM have been previously used to track animals in remote. Furthermore, whilst the Evans et al. \cite{Hol04} utilised UHF to transmit the location data, the findings and success of the GPS performance can be used as assurance as they also conducted their testing in Sabah, Malaysia. In addition, Evans et al. \cite{Hol04} have achieved a 58.1\% GPS success rate within the jungle, where there are more obstacles blocking the GPS signals. Our proposed system will be deployed in nearby towns and cities where the GPS signal should be stronger and easier to access.Ericsson et al. \cite{Hol04} have proposed a system that uses GPS/GSM collars to track moose in Sweden. This system was proposed in response to the problem of many previous techniques that used UHF (Ultra High Frequency) or VHF (Very High Frequency) radio collars, requiring significant effort in the field to collect data. The system proposed works by sending the location of the collar, which is attached to the moose, via SMS to a modem. The modem is connected to a database which stores the location data, which can then be viewed on an interactive map through a web application. After testing this system over a period of four years, they found it works and gives a new possibility to ‘supervise’ animals in almost real-time. They also found the biggest disadvantage was the limited or non-existing coverage of GSM networks in remote areas in the region. However, they also note that less GSM coverage is required for SMS transmissions in comparison to making a telephone call or GSM data transfer. In the second paper, Evans et al. \cite{Hol04} proposed to provide a method of tracking Malay Civets in Sabah, Malaysia, without using bulky collar units. The collars used in the proposed system utilise a GPS microchip, a 2300 or 2500 mAh battery, a UHF radio transmitter, a tri-axial accelerometer and an antenna. After testing the system, they found that the system returned a GPS success of 58.1\% across all tracked individuals over the testing period, up to 187 days long. Both of these works demonstrate that GPS and GSM have been previously used to track animals remotely. Furthermore, whilst Evans et al. \cite{Hol04} utilised UHF to transmit the location data, the findings and success of the GPS performance can be used as assurance as they also conducted their testing in Sabah, Malaysia. In addition, Evans et al. \cite{Hol04} have achieved a 58.1\% GPS success rate within the jungle, where more obstacles are blocking the GPS signals. Our proposed system will be deployed in nearby towns and cities where the GPS signal should be stronger and easier to access.

\section{Problem Definition}

As mentioned before, the motivation of this project is to enable the Danau Girang Field Centre in Sabah, Malaysia and the Sabah Wildlife Department to track the vehicles of poachers to understand the ecosystem of the gangs that fund and support them. Upon reviewing previous work in tracking vehicles and objects, we found that many of the proposed solutions were unsuitable for use by the DGFC \cite{PereraTR2018}. This is because many of the solutions utilise GPRS to communicate the location information, which is not viable for the DGFC. Whilst the focus of tracking the poachers would be in nearby towns where GPRS would be available, the system would be based in the DGFC site, which is in the middle of the Malaysian jungle where internet access is minimal. Therefore, a primary requirement of this project was not to depend on internet access.

Another important requirement was size. The tracking node should be small enough to be deployed into poachers’ vehicles through informants discreetly. Further, the locators used in this project had to be cheap and quick to configure and use. Because it is unlikely that we will be able to retrieve the nodes after deploying on poachers’ vehicles, they must also be battery-powered and maintain good accuracy and battery life. In order to extend the battery life, we also decided to add an extra battery pack. Finally, the last requirement was to connect the locators to a management software platform which allows tracking the movements of multiple nodes simultaneously. This software platform performs visualisation, configuration, and scheduling.

\section{System Prototype}

Our proposed system consists of three main components: (i) Locator, (ii) SMS modem, and (iii) Traccar tracking software. Let us briefly introduce each of the components.

\subsection{Locators}

\begin{figure*}[h!]
	\centering
	\includegraphics[scale = 1.0]{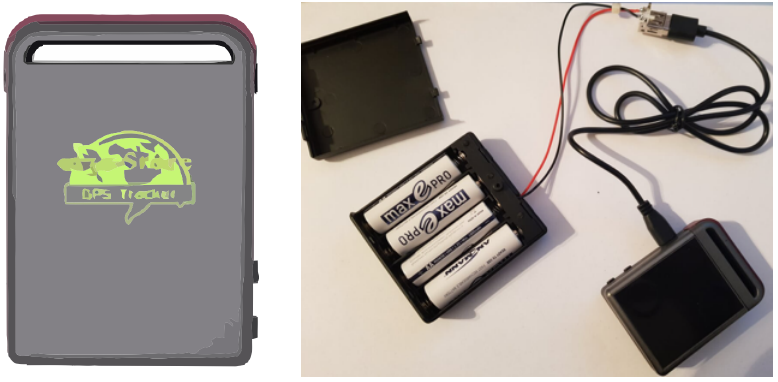}
	\caption{[Left] GPS Tracker Locator Device [Right] Extended battery pack}  
	\label{Fig:Locators}
\end{figure*}

The first and most important aspect of the system is the locators that will be attached to the vehicles. As already discussed, the main requirements for the locators are that they must be low-cost, battery-powered and utilise GSM to send the location data. The locator we chose is depicted in Figure \ref{Fig:Locators}. We also connected an additional battery pack to increase the lifetime of the tracker.

\begin{figure*}[h!]
	\centering
	\includegraphics[scale = 0.8]{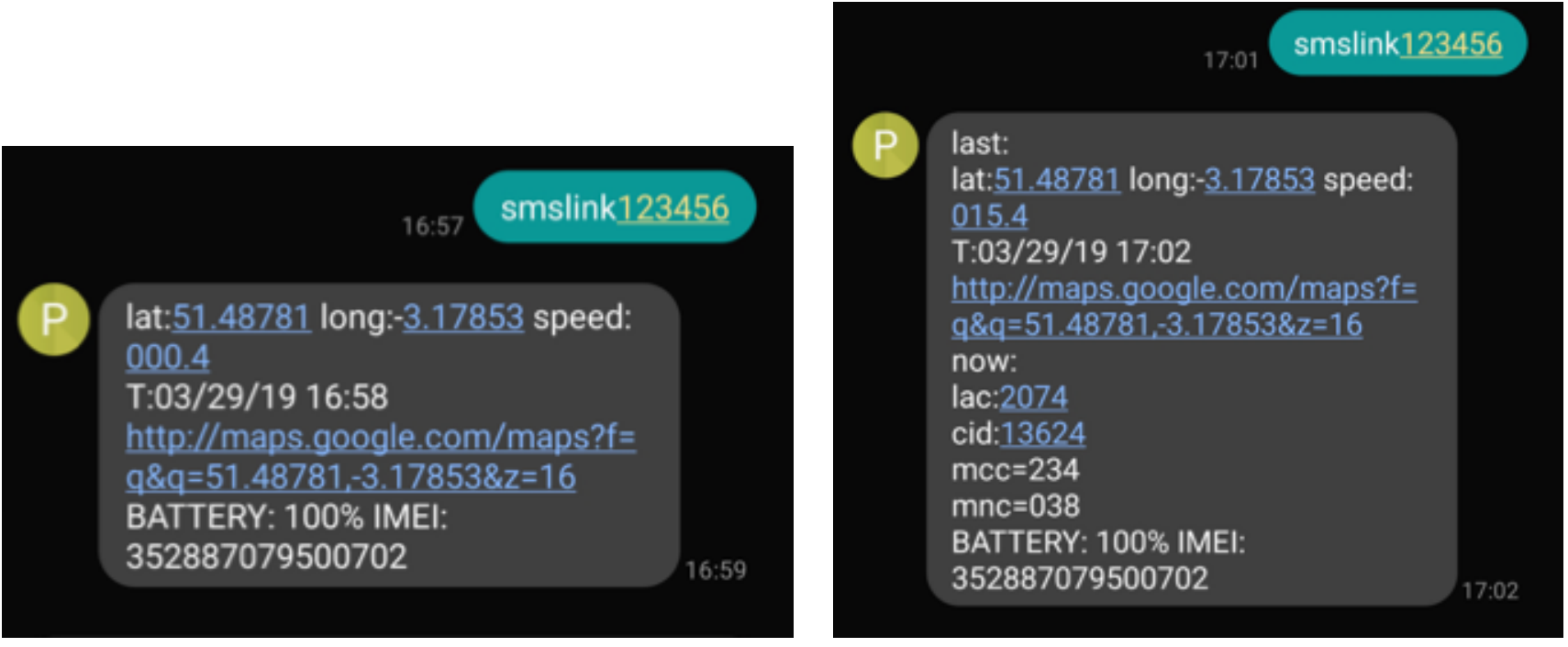}
	\caption{[Left] GPS Tracker Locator Device [Right] Extended battery pack}  
	\label{Fig:SMSMessage}
\end{figure*}

These devices retail for approximately £12.50 each. The locator is 64mm in length, 46mm in width and 17mm in height, making them small and compact enough to conceal easily. They come with standard plastic or magnetic lid that allows the user to covertly attach them to a car. Depending on the variation of locator, they also come with 2 batteries which are either 850 mAh or 1000 mAh (milliamp hour) in size. For this work, the 850 mAh variation was used for all of the locators. In order to work, the locators require a SIM card and operate by connecting to GPS through an antenna which is located at the top of the device. This allows the device to gain a location, which is then transmitted using SMS to the receiving number, as shown in Figure \ref{Fig:SMSMessage}(a). The SMS sent contains information such as the coordinates in Latitude and Longitude, the device's speed, a link to the location on Google Maps, the battery life of the locator as a percentage and the IMEI number. If the device cannot connect to GPS, it will return to its last location, as shown in Figure \ref{Fig:SMSMessage}(b).

This procedure of obtaining the location of the locator is activated by sending an SMS containing the message ‘smslink123456’ to the number of the device. However, we did not utilise the chosen locator device, providing extra functionality that could be useful in the future, such as a microphone and a TF memory card slot.

\subsection{GSM Modem}

\begin{figure*}[h!]
	\centering
	\includegraphics[scale = 1.0]{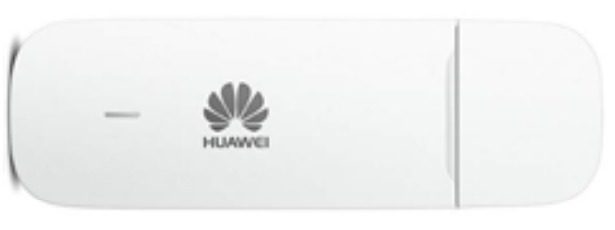}
	\caption{Huawei E3531 GSM Modem}  
	\label{Fig:HuaweiE3531GSMModem}
\end{figure*}

As mentioned in the above section, an SMS with the message ‘smslink123456’ must be sent to the locator over GSM to receive the location of the device. Therefore, a GSM modem was required in order to send SMS back and forth between the locators and the Software Platform. As depicted in Figure \ref{Fig:HuaweiE3531GSMModem}, we employed the Huawei E3531 GSM modem. It enables one to send and receive SMS messages and connect to the internet. Huawei E3531 GSM 3G modem costs approximately £20, making it relatively affordable in comparison to the modems’ more expensive 4G counterparts. Whilst the modem cost was one of the main reasons we opted for the Huawei E3531, other factors such as the modem being unlocked and providing an API, which we use in our system to send and receive messages, also influenced our decision to choose it.


\subsection{Traccar}

\begin{figure*}[h!]
	\centering
	\includegraphics[scale = 1.0]{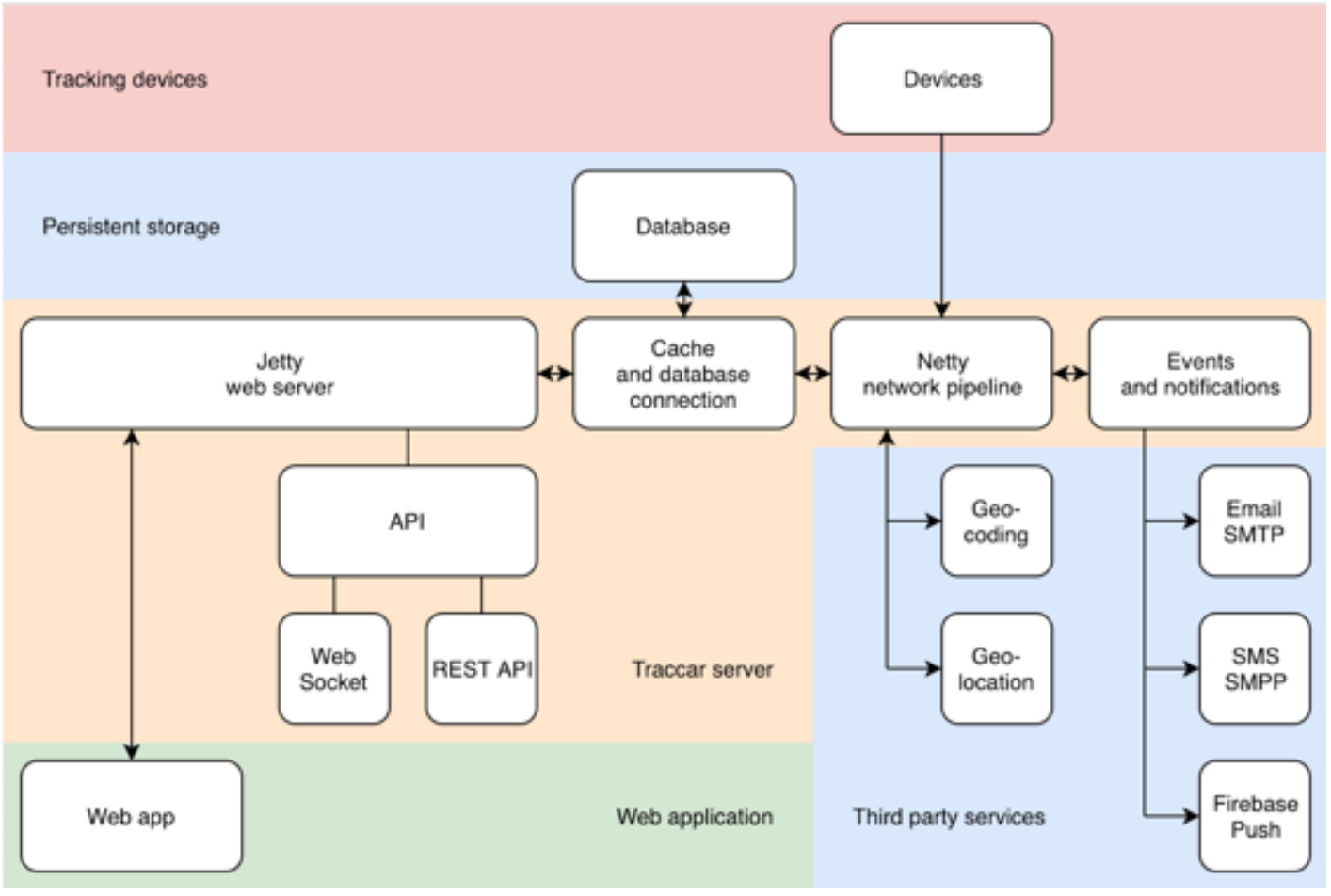}
	\caption{Traccar Server Architecture}  
	\label{Fig:TraccarServerArchitecture}
\end{figure*}

In order to visualise where the locators were moving around, we used Traccar (traccar.org), which is an open-source ‘modern GPS tracking system. More specifically, we used the Traccar server, which consists of a Java back-end and a Javascript frontend. By default, the back-end uses H2 databases. For this project, we configured it to use MYSQL. The Javascript frontend uses an embedded Jetty web server that is used to serve 2 components, the ‘Web API’ and the ‘Web Application’. The Web API consists of a REST API and Websocket, whilst the Web Application is based on the Sencha ExtJS framework and OpenLayers for map view. Traccar works by connecting to internet-connected GPS devices through a Netty network pipeline. The device communicates back to the server over the internet, and the server parses the information from the device and stores it in a database. This is then processed and displayed by the web application. The default process is demonstrated in the server architecture of Traccar in Figure \ref{Fig:TraccarServerArchitecture}.


Whereas for most situations, using Traccar with internet-connected devices would be perfectly viable, for this project, it is not. As already mentioned, this is because there is no infrastructure to communicate with the devices over the internet in the region where this project will be deployed. Therefore, for this project, the primary use of Traccar was the web application and the backend created by the Traccar server. The web application was modified and expanded to provide a visual method for communicating with the locators, providing a method for setting schedule scans and configuring locators for the installation, as well as visual and functional. Furthermore, the backend was used to add and modify databases to accommodate our changes to the Traccar web application. These modifications accommodated the changes made to the system whilst maintaining all of the original functions and uses of the Traccar server.  

As well as using Traccar for the backend which, as already mentioned, uses Java and JavaScript, we also used a Python Flask server whose purpose is to perform the various features we implemented into Traccar. Some of these features include contacting the SMS for individual and groups of locators, adding schedules and configuring devices for the installation. Predominantly the Flask server took variables from Traccar, processed them and then added them to the respective tables. In order to perform the functions that we implemented, the Flask server file uses 3 main Python packages- Flask, JSON and APScheduler. The Flask package enables the file to function as a server, and the JSON package is used to retrieve data back to Traccar, similar to the Traccar API. Moreover, the APScheduler package enables the scheduling feature to function, and we chose to use it as it provides a number of ways to run the scheduler as well as three different types of schedule: ‘date’, ‘interval’ and ‘cron’ making it incredibly versatile and useful. Utilising all three of these packages and a few others enables the file to perform its functions and mimic how the Traccar Web API works to provide consistency across the system.

\begin{figure*}[h!]
	\centering
	\includegraphics[scale = 0.45]{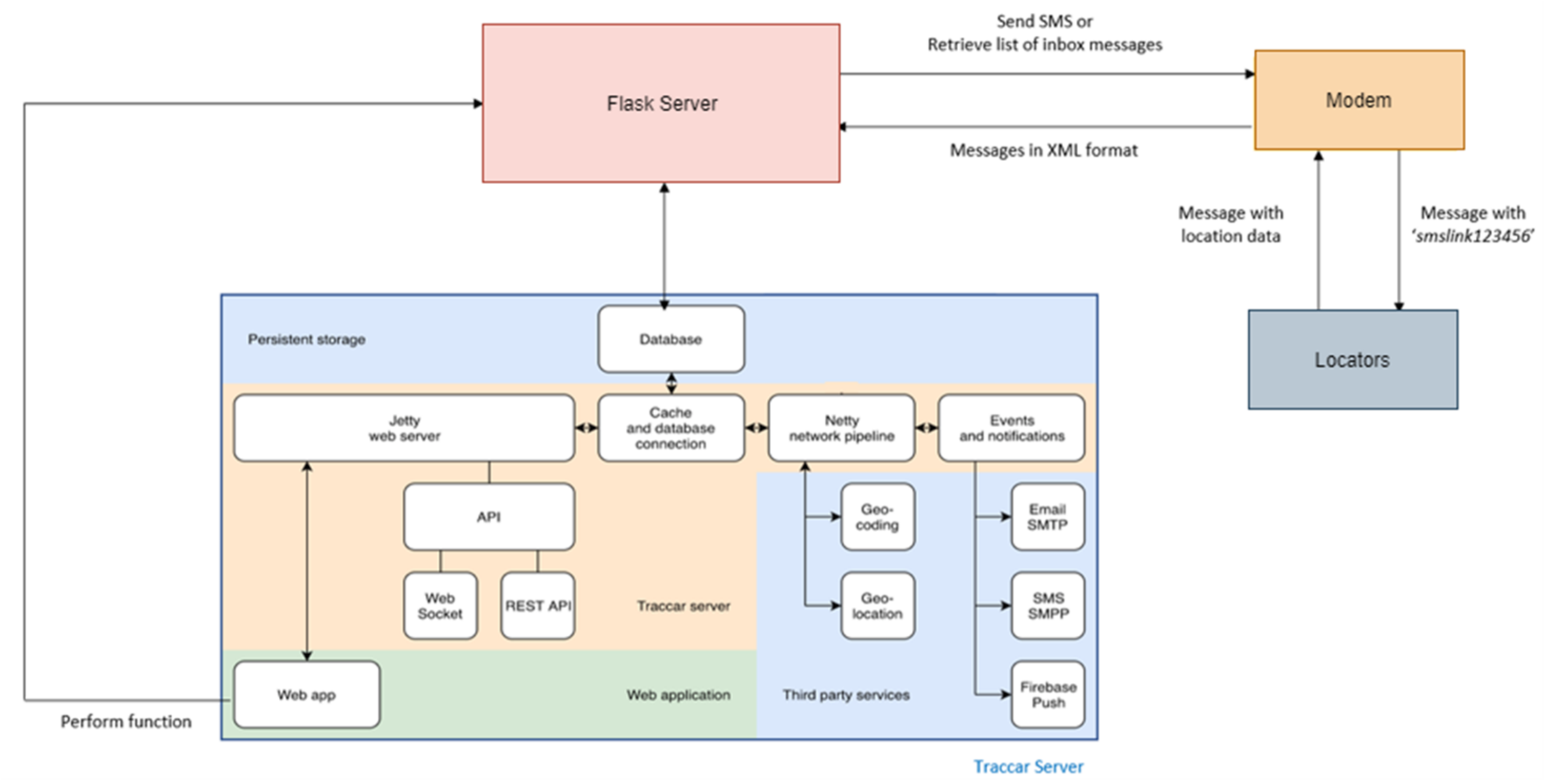}
	\caption{Proposed Extension to Traccar Architecture}  
	\label{Fig:ProposedExtensionToTraccarArchitecture}
\end{figure*}

In order to connect all the components of the system, we used the Flask server as the central component to interface with both Traccar, the modem and the locators. This works by directly interacting with the Traccar database to input data into the various tables required for the Flask server’s functions to run. These functions are coordinated by the Traccar web application that sends the relevant information to the Flask server, where it performs the function and outputs the data directly into the database. An example of this is when a new device is added to the installation through the web application, the data in the form fields are sent to the Flask server, where it is processed and then added to a table in the database. The data in this table is then returned to the web application when all the devices are listed. To connect to the modem, the Flask server does so through the modem API, which enables the system to send SMS and read the received messages. The Flask server uses this to send a message to the number of the locator, which then sends the location back as a response. The server retrieves the message, formats it and then inputs it into the Traccar database. Using the modem API allows the Flask server to interact with the modem and, therefore, the locators, making it a critical part of the system design. As you can see in Figure \ref{Fig:ProposedExtensionToTraccarArchitecture}, the result of using the Flask server as the central component is that the system architecture is clean and simple. It allows all the components to interface directly without relying on other unnecessary components, which is in stark contrast to the original system design idea for this project which can be seen in the GitHub repository.


\section{Evaluation}

We conducted performance evaluations in  (i) Cardiff, UK and conducted the feasibility testing (ii) in the wild. We are primarily reporting system performance evaluations conducted in Cardiff. We have omitted the details of feaibility testing deployment in Sabah, Malaysia, as they are sensitive in nature and could impact anti-poaching efforts.

\begin{figure*}[h!]
	\centering
	\includegraphics[scale = 0.30]{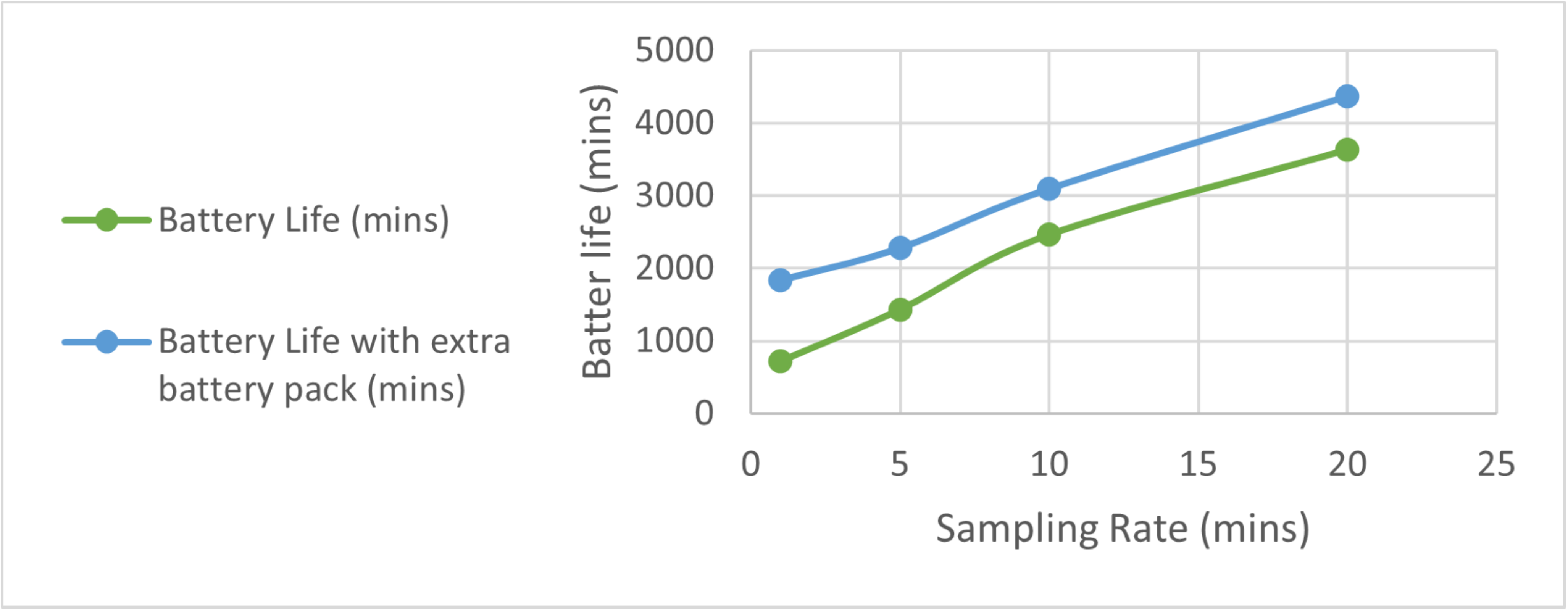}
	\caption{Battery Consumption over sampling rate (when continues monitoring)}  
	\label{Fig:Evaluations}
\end{figure*}

Upon testing the battery life of the locators across different location request intensities, it was found that the battery life lasts between 715 and 3637 minutes for highest intensity to lowest testing intensity, respectively. This translates to just under 12 hours for a location request every minute up to just over 60.5 hours for location requests every 20 minutes. It was also found from this testing that the battery life increases as the time between each SMS increases, as shown in Figure \ref{Fig:Evaluations}. In addition, response performance tests found that the average response time varies in between 30.5-42.8 seconds (i.e., time tracker node takes to get GPS coordinates and send the result back).


\section{Discussion And Lessons Learned}

\textbf{Sufficient Accuracy:} When in use and stationary, an average of 93.1\% of the responses are within 10 metres of the locator and 75.6\% of the responses are within 5 metres. This result demonstrates that the locators would be useful for tracking vehicles. The reason for this is that at a 10-metre accuracy, the user will still be able to determine the correct roads that the vehicle is travelling on to a very good level of confidence, and whilst this may be a bit more difficult for stationary vehicles, at a 5-10 metre accuracy level, it should allow the user to indicate the building or shop that the poacher is visiting. Even if the specific shop cannot be identified due to proximity or other factors, it will still give a general idea of the poacher’s interactions which is still useful intelligence. 

\textbf{Locators are good but not perfect:} The first evaluation we have been able to make on the system is that the locators we used are a viable solution for accurately tracking vehicles, as well as being cheaper than building a bespoke device using microcontrollers. For the most part, they have a good level of accuracy and perform well based on their cost versus the results achieved, and we would recommend using these locators in deployments where cost is a significant factor. 

\textbf{Not for Realtime Tracking:}  Having said that, however, it is worth noting that they are not perfect. We believe the long response times as well as the tendency for the locators to respond with error messages or incomplete messages, such as the ‘maps’ responses, make these locators unsuitable for very refined uses or deployments that require more intense real-time tracking, for example, every 30 seconds or less. 

\textbf{Extended Architecture and Scheduling feature enhanced Traccar:} Traccar is originally designed to work for internet-based tracking. However, we successfully extended it to support SMS-based tracking while enabling the rest of their features to be used for SMS-based tracking as well. We also developed a sophisticated scheduling algorithm (as well as a user interface to configure it) to facilitate a complex energy-efficient tracking schedule. Our extension allows us to query SMS tracker nodes based on different patterns (e.g., query only between 2 pm to 7 pm, weekends only, etc.) Such sophisticated scheduling functionality reduces the batter consumption drastically, allowing us to gather valuable information for a longer time.

\section{Conclusions And Future Work}

In this paper, we had a specific aim to achieve under specific constraints. Our overall objective was to develop an end-to-end hardware-software stack that allows conducting of tracking using SMS technology, enabling tracking studies in rural/remote areas without internet connectivity. In this study, we demonstrated the capability and suitability of the proposed SMS-based tracking systems through in-the-wild evaluations. We extended the internet-based tracking software platform Traccar to support SMS-based tracking in this work. We have provided the open-source code to this community to support tracking studies in areas where the internet is unavailable. One of our contributions was the scheduling feature we developed into the Traccar platform. Our scheduling feature allowed us to configure (i.e., sampling interval, activation period) any number of SMS-based tracker nodes individually or in groups. As a result, it allowed us to save energy and increase the lifespan of the SMS trackers. Through evaluation, we learnt that SMS tracker technology is suitable for tracking poachers without the internet. Further, the end-to-end hardware-software stack we enveloped is suitable for use in any domain requiring SMS-based tracking.

\section*{Acknowledgements}

We acknowledge the funding received by EPSRC (EP/T024372/1). The learning platform we developed is open source and available for the community: \url{https://gitlab.com/IOTGarage/tracking-poachers-using-gps-sms}
Demos is available: \url{https://youtu.be/lg5mumArEAU}. Code explanation is available: \url{https://youtu.be/0QuTXzMEJ3k}

\bibliographystyle{unsrt}  
\bibliography{library}
\end{document}